\documentstyle[11pt,epsf]{article}
\renewcommand{\figurename}{}\figurename

\renewcommand{\arraystretch}{1.3}
\catcode`\@=11
\def\marginnote#1{}

\newcount\hour
\newcount\minute
\newtoks\amorpm
\hour=\time\divide\hour by60
\minute=\time{\multiply\hour by60 \global\advance\minute by-\hour}
\edef\standardtime{{\ifnum\hour<12 \global\amorpm={am}%
        \else\global\amorpm={pm}\advance\hour by-12 \fi
        \ifnum\hour=0 \hour=12 \fi
        \number\hour:\ifnum\minute<10 0\fi\number\minute\the\amorpm}}
\edef\militarytime{\number\hour:\ifnum\minute<10 0\fi\number\minute}

\def\draftlabel#1{{\@bsphack\if@filesw {\let\thepage\relax
      \xdef\@gtempa{\write\@auxout{\string
          \newlabel{#1}{{\@currentlabel}{\thepage}}}}}\@gtempa \if@nobreak
    \ifvmode\nobreak\fi\fi\fi\@esphack} \gdef\@eqnlabel{#1}}
    \def\@eqnlabel{}
\def\@vacuum{}
\def\draftmarginnote#1{\marginpar{\raggedright\scriptsize\tt#1}}

\def\draft{
  \oddsidemargin -.5truein
  \def\@oddfoot{\footnotesize \sl preliminary draft \hfil
    \rm\thepage\hfil\sl\today\quad\militarytime}
  \let\@evenfoot\@oddfoot \overfullrule 3pt
    \let\label=\draftlabel
    \let\marginnote=\draftmarginnote
  \def\@eqnnum{(\theequation)\rlap{\kern\marginparsep\tt\@eqnlabel}%
    \global\let\@eqnlabel\@vacuum}

  }
\makeatletter
\newdimen\normalarrayskip              % skip between lines
\newdimen\minarrayskip                 % minimal skip between lines
\normalarrayskip\baselineskip
\minarrayskip\jot
\newif\ifold             \oldtrue            \def\new{\oldfalse}
\def\arraymode{\ifold\relax\else\displaystyle\fi} % mode of array entries
\def\eqnumphantom{\phantom{(\theequation)}}     % right phantom in eqnarray
\def\@arrayskip{\ifold\baselineskip\z@\lineskip\z@
     \else
     \baselineskip\minarrayskip\lineskip2\minarrayskip\fi}
\def\@arrayclassz{\ifcase \@lastchclass \@acolampacol \or
\@ampacol \or \or \or \@addamp \or
   \@acolampacol \or \@firstampfalse \@acol \fi
\edef\@preamble{\@preamble
  \ifcase \@chnum
     \hfil$\relax\arraymode\@sharp$\hfil
     \or $\relax\arraymode\@sharp$\hfil
     \or \hfil$\relax\arraymode\@sharp$\fi}}
\def\@array[#1]#2{\setbox\@arstrutbox=\hbox{\vrule
     height\arraystretch \ht\strutbox
     depth\arraystretch \dp\strutbox
     width\z@}\@mkpream{#2}\edef\@preamble{\halign
\noexpand\@halignto
\bgroup \tabskip\z@ \@arstrut \@preamble \tabskip\z@ \cr}%
\let\@startpbox\@@startpbox \let\@endpbox\@@endpbox
  \if #1t\vtop \else \if#1b\vbox \else \vcenter \fi\fi
  \bgroup \let\par\relax
  \let\@sharp##\let\protect\relax
  \@arrayskip\@preamble}
\def\eqnarray{\stepcounter{equation}%
              \let\@currentlabel=\theequation
              \global\@eqnswtrue
              \global\@eqcnt\z@
              \tabskip\@centering
              \let\\=\@eqncr
 \halign to \displaywidth\bgroup
    \eqnumphantom\@eqnsel\hskip\@centering
    $\displaystyle \tabskip\z@ {##}$%
    \global\@eqcnt\@ne \hskip 2\arraycolsep
         $\displaystyle\arraymode{##}$\hfil
    \global\@eqcnt\tw@ \hskip 2\arraycolsep
         $\displaystyle\tabskip\z@{##}$\hfil
         \tabskip\@centering
    &{##}\tabskip\z@\cr}
\begingroup\ifx\undefined\newsymbol \else\def\input#1 {\endgroup}\fi
\newfont{\hr}{msbm10}
\newfont{\ams}{msam10}
\hoffset= - 1.0in         % switch off for draft style
\def\beq{\begin{equation}}
\def\eeq{\end{equation}}
\def\ba{\beq\new\begin{array}{c}}
\def\ea{\end{array}\eeq}
\def\be{\ba}
\def\ee{\ea}

\def\N2{${\cal N}=2$}

\def\1N{${\cal N}=1$}
\def\4N{${\cal N}=4$}
\def\nn{\nonumber}

\title{{\bf
On the Need for Phenomenological Theory of $P$-Vortices\\
or\\
Does Spaghetti Confinement Pattern Admit Condensed-Matter Analogies?}
\vspace{.5cm}}
\author{{\bf A.Mironov}\thanks{E-mail:
\ mironov@itep.ru; mironov@lpi.ac.ru}
\date{ } \\
{\small {\it Theory Department, Lebedev Physics Institute}
and {\it ITEP, Moscow, Russia}}\\
{\bf A.Morozov}\thanks{E-mail: \ morozov@itep.ru}
\date{ } \\ {\small
{\it ITEP, Moscow}
}\\
{\bf T.N.Tomaras}\thanks{E-mail:
\ tomaras@physics.uoc.gr}
\date{ } \\
{\small {\it Department of Physics and Institute of Plasma Physics, University of Crete,
and Fo.R.T.H., Greece}}}

\begin{document}

\setcounter{footnote}{3}

\maketitle

\vspace{-15cm}

\begin{center}
\hfill FIAN/TD-12/04\\
\hfill ITEP/TH-45/04\\
\end{center}

\vspace{13.5cm}

\begin{abstract}
Usually the intuition from condensed-matter physics is used to
provide ideas for possible confinement mechanisms in gauge
theories.
Today, with a clear but puzzling ``spaghetti'' confinement pattern,
arising after a decade of lattice computer experiments,
which implies formation of a fluctuating net of peculiar magnetic vortices
rather than condensation of the homogeneously distributed magnetic monopoles,
the time is coming to reverse the logic and search for similar
patterns in condensed matter systems.
The main thing to look for in a condensed matter setup
is the simultaneous existence of narrow
tubes ($P$-vortices or 1-branes) of direction-changing electric
field and broader tubes (Abrikosov lines) of magnetic field,
a pattern dual to the one,
presumably underlying confinement in gluodynamics.
As a possible place for this search we suggest systems
with coexisting charge-density waves and superconductivity.
\end{abstract}

\vspace{1cm}

\section{Introduction}

A possible resolution of the confinement problem \cite{conf}-\cite{meanf}
should answer questions
at two related but somewhat different levels:\footnote{
We discuss confinement as a pure gluodynamical problem and ignore all
issues related to fermion condensates and chiral symmetry breaking.
In the real-world QCD, the effects related to
light quarks, can be more important for a large part of hadron
physics and even the {\it dominant} confinement mechanism may be
different \cite{Gribovlq}. Because of this, in the study of confinement
in {\it gluodynamics} one should rely more upon computer than
accelerator experiments.

We also do not dwell upon the promissing ``holistic'' approaches to
confinement, exploiting various general properties of gluodynamics
\cite{genpro} or building one or another kind of self-consistent
approximation to correlation functions \cite{pertco}, \cite{meanf}.
Instead we discuss the lattice-experiment results, providing a
{\it microscopic} description of relevant field configurations and their
common properties and wonder if this mysterious pattern was ever observed
in other types of physical systems.
}

(i) It should allow for a reliable evaluation of various quantities,
such as the gap in the spectrum of perturbations around the {\it true
vacuum}, the string tensions in the area laws for the Wilson loops
in different representations, as well as the masses of glueballs
and other hadrons (when light quarks are taken into consideration).

(ii) It should provide a simple qualitative ``picture'' of how the
vacuum is formed, how does the linear potential arise between remote
sources with non-vanishing $N$-alities in the absence of
light quarks and how are the massive colorless hadrons formed
in the absence as well as in the presence of light quarks.

Of principal importance for developement of theoretical
(not computer-experimental) quantitative methods at level (i)
would be identification of the  true vacuum $|vac \rangle$ --
a functional of fields at a given moment of time, which is the lowest
eigenstate of the non-perturbative Yang-Mills Hamiltonian,--
with all the other eigenstates presumably separated from
$|vac \rangle$ by a non-vanishing gap.

The relevant approach to (ii) would rather identify a relatively
small subspace in the space of all field configurations (labeled
by a sort of collective coordinates) and substitute the original
problem of Yang-Mills dynamics by that of a more or less familiar
{\it medium} -- QCD aether (like a gas of monopoles or $P$-vortices,
a dual superconductor or something else).
The underlying belief here is that the original
functional integral at low energies gets dominant contribution from
a restricted set of field configurations, and thus
can be substituted by some more familiar effective theory,
describing -- at least qualitatively -- the low-energy quantities
as averages over this auxiliary {\it medium} and expressing the problems
of low-energy quantum Yang-Mills theory through those of the {\it medium}
dynamics.

{\it Understanding} of confinement requires certain achievements at
both levels (i) and (ii): the existence of a ``picture'' is the thing that
distinguishes ``understanding'' from just ``calculability'', while the
possibility to make calculations or at least estimates is a
criterium for selection of a correct ``picture'' among the alternative ones.
%\footnote{{\bf version:} since both quantitative results and qualitative
%"visualization" in terms of some familiar pattern are needed, and neither
%one directly implies another. }

%Confinement mechanism should imply, that:

%($\alpha$) force lines from sources with non-vanishing $N$-ality
%(i.e. in specific representations of the gauge group)
%can not terminate;

%($\beta$) the {\it medium} repels force lines.

%These two properties are believed to imply that the force lines
%between point-like
%sources should collimate to form narrow tubes and thus give rise to linear
%attractive potentials.

The problem of confinement consists of two parts: one should explain, why

{\bf ($\alpha$)} all gauge fields are screened (i.e. all gluons,
electric and magnetic, acquire effective masses $\sim \Lambda_{QCD}$) and

{\bf ($\beta$)} still, there is a peculiar {\it long-range}
color-electric interaction, described by a narrow tube, where electric
force lines (carrying a flux with non-vanishing $N$-ality, i.e. in
representation, which can not be obtained in a product of adjoints, so that
the tube is stable against string-breaking, caused by creation of a set of
gluons) are collimated and give rise to the
linear interaction potential $V(R) \sim \sigma R$ at $R \gg
\Lambda_{QCD}^{-1}$, with the string tension $\sigma \sim \Lambda^2_{QCD}$ and
the string width $r_e \sim \Lambda_{QCD}^{-1}\log (R\Lambda_{QCD})$.

We call this double-face situation the dual Meissner-Abrikosov (MA) effect.

The spaghetti vacuum pattern \cite{orNO}, to be discussed below, implies that
in addition to ($\alpha$) and ($\beta$),

{\bf ($\gamma$)}
one more {\it long-range} interaction survives, described by
a {\it very} narrow tube ($P$-vortex or $1$-brane),
with collimated color-magnetic force lines, populated by $0$-branes,
looking in certain aspects like magnetic monopoles and antimonopoles,
with the direction of the magnetic field reversed at the locations of the
$0$-branes,

{\bf ($\delta$)}
the $P$-vortices can merge and split, they form a dense net,
percolating through the whole volume.

Thus in some sense the dual MA effect is complemented by a kind of
ordinary MA effect, though magnetic Abrikosov tubes carry a good
deal of additional structure (moreover, as we will discuss below,
the oversimplified description
of this structure, as given in ($\gamma$) is not gauge-invariant and
thus is not fully adequate).

\section{Screening in Abelian theory}

It is well known that the MA effect {\it per se} does not require
a non-Abelian gauge theory for its manifestation. It can be
discussed already at the Abelian level.

There are many ways to obtain one or another kind of the screening effect
($\alpha$) and many of them allow for one or another kind of long-range
interactions to survive.

\paragraph{Massive photon.}
Complete screening with no long-range interactions is described
by the effective Lagrangian of the type
\be
\frac{1}{e^2}F_{\mu\nu}^2 + m^2A_\mu^2.
\ee
It explicitly breaks gauge invariance and contains non-propagating
degrees of freedom $A_0$, giving rise to instanteneous, but still
screened, interaction.

\paragraph{Debye screening.}
It appears in ordinary conductors, electrolytes and some phases of plasma
and is described by the effective Largangian
\be
\frac{1}{e^2}F_{\mu\nu}^2 - E_i\frac{m^2}{\vec\partial^2}E_i.
\ee
It explicitly breaks Lorentz invariance and completely screens static
electric fields, while magnetic and time-oscillating electric fields remain
long-range. The massive term is usually produced by the process shown in
Figure 1 and $m^2$ is proportinal to the concentration $n_0$ of electric
charges in the medium. If these charges are not originally present,
then $m^2 \sim n_0$ is either due to non-vanishing temperature, or,
if the temperature is zero, to the probability of charge-anticharge creation
by an imposed external electric field. This probability and thus $m^2$
normally contains extra powers of 4-momenta, so that the screening mechanism
gets essentially softened and leads, for example, to the slow
{\bf running coupling} phenomenon in QED and QCD, described (in these
Lorentz-invariant cases) by the effective Lagrangian
\be
F_{\mu\nu}\frac{1}{e^2(\Delta)}F_{\mu\nu}.
\label{runco}
\ee
In $3+1$ dimensions the $\Delta$-dependence is just logarithmic,
at least in the leading approximation,
so that no real screening takes place, gauge fields remain massless.
In non-Abelian theories magnetic interactions also enter the game,
producing the anti-screening effect in (\ref{runco}),
overweighting the screening one \cite{hughes}. It is not quite clear whether just
this anti-screening could lead to the confinment effect
when moving beyond the leading logarithm approximation (see, e.g., \cite{pertco}).

%\bigskip

%\hrule height 0.4mm

\bigskip

\paragraph{\large {\bf Figure 1.}}
{\footnotesize
The origin of the gauge-field mass in Debye-screening mechanism:}

\bigskip

\setcounter{figure}{0}
\def\thefigure{\alph{figure})}

\begin{figure}[h]
\epsfxsize 300pt
\begin{center}
\epsffile{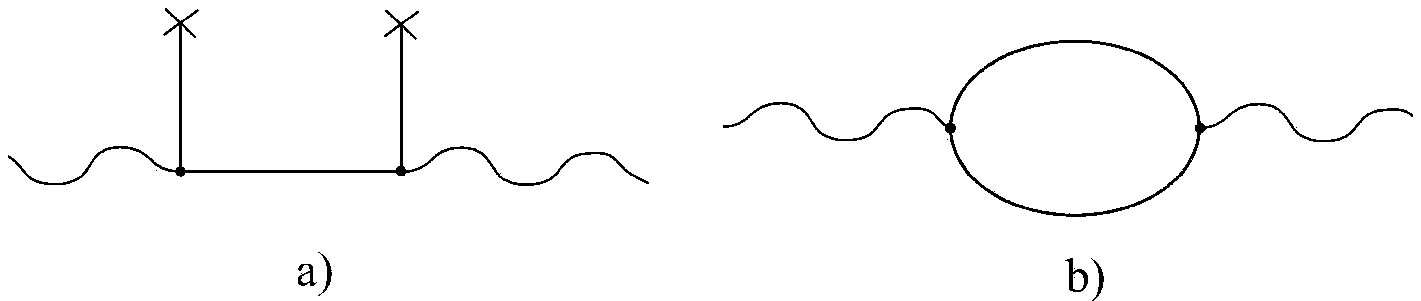}
\end{center}
\caption{\footnotesize
The case when charged particles are originally present in the medium.
The entire diagram is proportional to the concentration $n_0$ of
these particles in the medium. For non-vanishing temperatures (unavoidable
in any lattice calculations) $n_0$ is never zero (but can be exponentially
small).}
\caption{\footnotesize
The case when the charged pairs are created in the medium (including
the physical vacuum) by the gauge field itself. In this case
the screening is usually
much softer and can result in a slow running of a
coupling constant rather than in exponential screening.}
\end{figure}

%\bigskip

%\hrule height 0.4mm

\bigskip

To be more precise, in realistic systems the effective Lagrangian
(in the case of linear response, i.e. weak fields) is
expressed in terms of the dielectric constant
\footnote{
Note that the formulation in terms of dielectric constant and magnetic
permeability $\mu$ can be useful in the search for solid-state counterparts
of the confinement phenomenon (see e.g. \cite{Kir}): the electric confinement
(like that in QCD) can be described by $\epsilon=0$, while the magnetic
confinement (like the Meissner effect in superconductors) is attributed
to $\mu=0$.}
$\epsilon_{ij}
\equiv\left(\delta_{ij}-{k_ik_j\over \vec k^2}\right)
\epsilon_{\bot}(\omega,\vec k)+{k_ik_j\over\vec k^2}
\epsilon_{\|}(\omega,\vec k)$:
\be
{\cal L}=\frac{1}{e^2}\left[F_{\mu\nu}^2 + (\epsilon_{\bot}-1)\vec E^2
+(\epsilon_{\bot}+\epsilon_{\|})\left(\hbox{div}\vec E {1\over\bf
\vec\partial ^2} \hbox{div}\vec E\right)
\right]
\ee
and is not universal, since the frequency and momentum dependence of
$\epsilon_{\|}$ and $\epsilon_{\bot}$ can be very different in different
regimes. Important for the Debye screening (long-distance exponential decay
of the field correlator) is the presence of a singularity in the
longitudinal dielectric constant at large distances (small $\vec k^2$)
\cite{LL}:
$\epsilon_{\|}=1+{e^2m^2\over\vec k^2}+{\cal O}(\omega)$, where the omitted
terms describe highly non-trivial frequency dependence. Indeed, the
static correlator is
\be\label{ecor}
\left< E_iE_j\right>\sim{k_ik_j\over \epsilon_{\|} \vec k^2}=
{k_ik_j\over \vec k^2 +P_{00}}
\ee
where $P_{00}= (\epsilon_{\|}-1)\vec k^2$ is the static value of the
component of the photon polarization operator $P_{\mu\nu}$ ("electric" mass
\cite{GPY}).

\paragraph{Dual Debye screening.}
It would be described by a dual effective Largangian of the type
\be
\frac{1}{e^2}F_{\mu\nu}^2 + H_{i}\frac{m^2}{\vec\partial^2}H_{i}
\ee
and imply screening of static magnetic fields.
It is unclear if any condensed-matter systems with this type of
behaviour have been already discovered. In ordinary electrodynamics
without magnetic charges, we have a counterpart of (\ref{ecor}):
\be
\left<H_iH_j\right>={\vec k^2\delta_{ij}-k_ik_j\over \vec k^2 +P}
\ee
where the ``magnetic'' mass $P$ is given by the static value of the spatial
components of the photon polarization operator ($P_{ij}
\stackrel{\omega=0}{=}\left(
\delta_{ij}-{k_ik_j\over \vec k^2} \right)P$ due to the gauge invariance).
In a gas of magnetic monopoles it becomes (see Polyakov's book in
\cite{conf})
\be
\left<H_iH_j\right>=\delta_{ij}-{k_ik_j\over k^2+M^2}
\ee

\paragraph{Chern-Simons screening.}
It is described by the peculiar gauge invariant Lagrangian,
\be
\frac{1}{e^2}F_{\mu\nu}^2 +
m^{\alpha\ldots\beta}\epsilon_{\mu\nu\lambda\alpha\ldots\beta}
A_\lambda F_{\mu\nu}.
\ee
It describes aspects of the Hall effect and related phenomena,
is Lorentz invariant ($m$ is a scalar)
only in $2+1$ dimensions, and -- only in this dimension --
makes the photon massive, but still the {\it long-range}
Aharonov-Bohm interaction survives \cite{Kog}.

\paragraph{Abelian Higgs model.}
The ordinary (not the dual) Meissner-Abrikosov effect is modelled by
the Abelian Higgs (Landau-Ginzburg) effective Lagrangian
\be\label{ah}
\frac{1}{e^2}F_{\mu\nu}^2 +
|D_\mu\phi|^2 + \lambda (|\phi|^2-m^2)^2.
\ee
After $\phi$ condenses, $\langle \phi \rangle\ = me^{i\theta}$,
the gauge fields become massive, thus giving rise to effect
($\alpha$): the Meissner effect for magnetic and electric fields.
However, actually the mass is acquired not by $A_\mu$ field, but
rather by a gauge invariant combination
$\hat A_\mu = A_\mu - \partial_\mu\theta$, thus the mode
$\hat A_\mu = 0$ can still propagate through large distances,
and this explains the effect ($\beta$): emergence of Abrikosov tubes.
$\hat A_\mu=0$ does not imply that $A_\mu = \partial_\mu\theta$ is
pure gauge, if $\theta(x)$ is singular and $\oint_C A_\mu dx^\mu \neq
0$ for some contours $C$. In an Abrikosov tube stretched along the $z$ axis
$\theta = \arctan \frac{y}{x}$ is the angle in the $xy$ plane and
$C$ is any contour in this plane, encircling the origin. Since
$\theta$ is the phase of the smooth field $\phi$, the modulus
$|\phi|$ should vanish on the $z$ axis, where $\theta$ is not well defined,
i.e. the condition $|\langle \phi \rangle | = m$ is destroyed in the
vicinity of $z$ axis, in a tube with the cros-section $\Sigma = \pi r_m^2$.
This causes the energy $\lambda m^4\Sigma$ per unit length of the tube,
while the energy of magnetic flux $\Phi$ in the tube is $\sim
\left(\frac{\Phi}{\Sigma}\right)^2\Sigma = \frac{\Phi^2}{\Sigma}$.
Minimization of the sum of these terms with respect to $\Sigma$ defines the
characteristic width of the tube $\Sigma_m = \pi r_m^2 \sim
\frac{\Phi}{\sqrt{\lambda}a^2}$.

If electric charges $q$ smaller than that of the Higgs field $\phi$
are present in the theory, then $q\Phi$ can be smaller than $1$ and
Aharonov-Bohm effect will be observed when such charges travel around
the Abrikosov tube at any distance: thus, even though all gauge fields are
massive, the Aharonov-Bohm interaction also remains long-range (unscreened)
\cite{Ba}.

The technical reason allowing {\it magnetic} Abrikosov lines to exist is that
the equation $F_{xy} = \delta(x)\delta(y)$ can be easily resolved:
$A_x = \partial_x \arctan\frac{y}{x}, \
A_y = \partial_y \arctan\frac{y}{x}$ and the Higgs field just provides a
source of the needed form, with {\it electric} current
$J_x = \partial_y F_{xy} = \delta(x)\delta'(y), \
J_y = -\partial_x F_{xy} = -\delta'(x)\delta(y)$
rotating around the $z$-axis.

In order to obtain an {\it electric} Abrikosov line one would need to solve the
equation $F_{0z} = \delta(x)\delta(y)$, which violates Bianchi identity
and requires the existence of a {\it magnetic} current (rotating around
the $z$-axis) and thus, in a Lorentz invariant setting, of magnetic charges
(monopoles).\footnote{
Similarly, in order to have a {\it magnetic} tube, where the field is
not constant along the line, in particular it changes direction at some
points $z_a$, one needs
to solve an equation $F_{xy} ={1\over 2}
\delta(x)\delta(y)\prod_a {\hbox{sign}}(z-z_a)$
which violates Bianchi identity at $x=y=0,\ z=z_a$ and thus requires
magnetic charges (monopoles) at these points.
}
Thus, in order to describe confinement with properties ($\alpha$) and
($\beta$), where the {\it dual} MA effect is needed,
one often makes use of the {\bf dual Abelian
Higgs model} (the dual superconductor model),
where the Higgs field $\tilde\phi$ is magnetically charged,
i.e. interacts with the dual field $\tilde A_\mu$, such that
$\tilde F_{\mu\nu} = \partial_\mu\tilde A_\nu - \partial_\nu\tilde A_\mu =
\frac{1}{2}\epsilon_{\mu\nu\alpha\beta}F_{\alpha\beta} =
\epsilon_{\mu\nu\alpha\beta}\partial_\alpha A_\beta$.
In this type of scenarios the role of non-Abelian degrees of freedom is
thought to be the imitation of Higgs degrees of freedom
%\footnote{The idea
%is to fix an Abelian gauge which is specified by diagonalizing an adjoint scalar.
%The remnant $U(1)^{\kappa}$ ($kappa$ being the rank of the gauge group)
%gauge invariance then acts on the non-diagonal gauge fields as on even charged
%scalars so that one could hope monopoles in such theory exist.
%}
(see, for example,
$W^\pm$ in eq.(\ref{SU2MAP}) below and ref.\cite{FN})
and the problem is to find a mechanism,
leading to their appropriate condensation.

As already mentioned, the lattice experiments (see Section 4 below) imply that
the real pattern (and, perhaps, the mechanism) of confinement can be
more sophisticated and may imply coexistence of ($\beta$) electric and
($\gamma$) structured magnetic tubes.
Therefore, it is important to note that no Abelian model is known,
which would allow the coexistence of magnetic and electric MA effects, e.g.
no effective Lagrangian of the form
\be
\frac{1}{e^2}F_{\mu\nu}^2 + m_m^2\hat A_\mu^2 + m_e^2\hat{\tilde A_\mu}^2
\ee
is allowed. Therefore, if such coexistence is not an artifact of
lattice experiments (what is not considered too probable nowadays),
it requires construction of more sophisticated models. A natural hope is
that such models can be straightforwardly built in modern string theory
(involving branes) and realized in condensed-matter systems.

Note that what is needed is some kind of restoration, at least partial, of
electro-magnetic duality, present in Abelian photodynamics. This duality
is usually broken by all known relevant modifications: by the introduction of
electric charges (without adding their magnetic counterparts), by
embedding into non-Abelian theory (where electric and magnetic interactions
of gluons are different), by the addition of a Chern-Simons term, or by
coupling to Higgs scalars and going to a superconducting phase.
Lattice experiments strongly suggest the need for some -- yet unstudied --
(topological, i.e. with the field-content of field, not string, theory)
stringy phases with both ``fundamental'' and $D1$ strings
present, where screening and MA phenomena do not
contradict electromagnetic duality.

\section{$3d$ compact QED}

The sample example \cite{Pol} of confinement proof
in Abelian $2+1$ dimensional
compact electrodynamics (embedded into the non-Abelian
Georgi-Glashow model to justify compactness and
provide ultraviolet regularization, rendering finite the instanton
action), deals actually with {\it random confinement} \cite{orNO,meanf}
and with {\it Wilson's confinement criterion} \cite{conf}:
no fluxes acquire vacuum averages, only their squares,
$\langle \Phi \rangle = 0$, $\langle \Phi^2 \rangle \neq 0$,
and this is enough to provide the area-law behaviour for the
Wilson-loop averages.
In this example the relevant {\it medium} in 2 space dimensions
is obtained as a time-slice of an instanton gas with Debye screening.
Instantons in Abelian $2+1$-dimensional theory are just
ordinary $3$-dimensional monopoles and antimonopoles with magnetic fields
\be
H_\mu = \epsilon_{\mu\nu\lambda} F^{\nu\lambda} =
\pm g\frac{r_\mu}{r^3},
\ee
or rather
\be
H_\mu = \pm g\frac{r_\mu}{(r^2 + \varepsilon^2)}e^{-r/\xi},
\ee
where $\epsilon$ and $\xi$ provide ultraviolet (from the underlying
non-Abelian theory) and infrared (from the Debye screening in the
monopole-\-antimonopole gas) regularizations respectively; $g$ is the
monopole charge, normalized so that $2eg=$integer.
Thus the {\it medium}
looks like a set of appearing and disappearing vortex-antivortex
pairs with the pseudoscalar $2d$ magnetic and vector
$2d$ electric fields
\be
B = \epsilon_{ij} F^{ij} = \pm \frac{gt}{(\vec x^2 + t^2)^{3/2}},
\nn \\
E_i = F_{0i} =
\pm g\frac{\epsilon_{ij}x^j}{(\vec x^2 + t^2)^{3/2}}.
\ee
The field $E_i$, produced by the time-variation of $B$, has
non-trivial vorticity and thus contributes to the rectangular Wilson
average over this medium
\be
\langle \exp \left(ie\oint_C (A_0 dt + A_i dx^i)\right) \rangle \ =
\ \langle \exp \left(ie \int_S \vec E \cdot d\vec x dt\right) \rangle
\ee
where the contour $C$ lies in the $xt$ plane and $S$ intersects the
$xy$ plane by a segment $\tilde C$. The contribution of a vortex
to the integral $\int_{\tilde C} \vec E \cdot d\vec x$ is equal to
\be
\pm \int_{-L}^L \frac{ydx}{(x^2 + y^2 + t^2)^{3/2}}
\sim \pm\frac{2y}{y^2+t^2}
\ee
for $L \gg \sqrt{y^2 + t^2}$ (with the distance $\sqrt{y^2+t^2}$ actually
bounded from above by the Debye radius $\xi$) and further integration
over $t$ gives for a contribution of a vortex:
\be
\pm 4\pi g\Phi = \pm 2\pi g\frac{y}{|y|} = \pm 2\pi g{\rm sign} (y),
\ee
provided the vortex lies in a slice of width $\xi\ll L$ around the
surface $S$. This flux is one-half of the full flux $4\pi g$ of the
charge-$g$ monopole.
The factor $1/2$ appears here because only half of
the vorticity of $\vec E$ contributes to the integral.
Since contributions of vortices and antivortices have
opposite signs, the average of $\int \vec E d\vec x dt$ itself
is of course vanishing, but the even powers of this integral, and
thus the Wilson exponent, can have non-vanishing averages.
The simplest estimate with the help of Poisson
distributions gives \cite{HT}:
\be
\langle \exp \left(ie\oint_C (A_0 dt + A_i dx^i)\right) \rangle \ =
\nn \\ =
\sum_{n_+,n_- = 0}^\infty
\left[e^{-\bar n}\frac{\bar n^{n_+}}{n_+!}\right]
\left[e^{-\bar n}\frac{\bar n^{n_-}}{n_-!}\right]
e^{4\pi i eg(n_+-n_-)\Phi} =
e^{-2\bar n(1-\cos(4\pi eg\Phi))}
\label{na3co}
\ee
Since the average number of contributing vortices and antivortices
is  $\bar n = \xi A_S n_0$, where $A_S$ is the area of the surface $S$
and $n_0$ is the concentration of vortices (depending, primarily on the
instanton action, which in turn is defined by the ultraviolet
regularization), one obtains the area law for the Wilson loop, at least for
the minimal value $eg=1/2$ allowed by the Dirac quantization
condition
\footnote{
There are corrections to this oversimplified calculation
\cite{AG,HT}, which in particular can destroy the prediction of (\ref{na3co}),
that confinement disappears
for {\it even} magnetic charges (when the relevant flux $\Phi$ is integer).
}.
Similarly, one could calculate the average
\be
\langle \exp \left(ie\oint_C (A_0 dt + A_i dx^i)\right) \rangle \ =
\ \langle \exp \left(ie \int_S B d x dy\right) \rangle
\ee
of the space-like Wilson loop, with $S$ lying in the $xy$-plane and bounded
by the curve $C$. This average is given by the same formula (\ref{na3co}).

Another {\it interpretation} of the same
calculation \cite{CGD}, implies that the distribution of vortices is
{\it affected} by the presence of the loop, so that the vortices and
antivortices get concentrated around the surface $S$ and {\it screen} it.

\section{Confinement in $4d$}

In $3+1$ dimensions no such simple calculation from first principles is yet
known. The main difference is that ordinary instantons in $3+1$ dimensions
are no longer charged: their field vanishes too fast at infinity and,
therefore, the confinement mechanism should involve an additional
{\it dissociation} of instantons into something like magnetically
charged {\it merons} \cite{CGD,mer}.
Time slices of instantons are now $3d$ objects, namely monopole-antimonopole
pairs (if looked at in a special gauge),
and the instanton describes the process of their spontaneous creation
and annihilation.

\paragraph{\large {\bf Figure 2.}}

{\footnotesize Possible phases of the recombinant plasma of instanton gas.}

\bigskip
\setcounter{figure}{0}
\def\thefigure{\alph{figure})}

\begin{figure}[h]
\epsfxsize 350pt
\epsffile{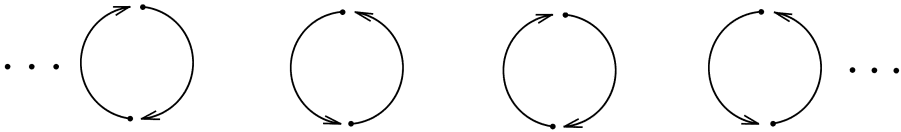}
\caption{\footnotesize
{\it Recombinant phase} (ordinary instanton gas in $3+1$ dimensions):
each instanton is the
process of creation and annihilation of a monopole-antimonopole pair.}
\end{figure}

\begin{figure}[h]
\epsfxsize 350pt
\epsffile{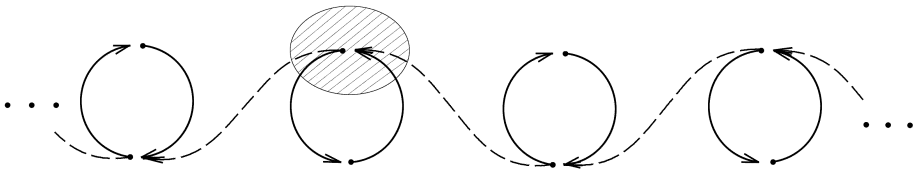}
\caption{\footnotesize Transition to
{\it jumping recombinant phase} (instantons dissociate into merons):
created pairs do not coincide with annihilating pairs. The shadowed domain
corresponds to the meron.}
\end{figure}

\begin{figure}[h]
\epsfxsize 350pt
\epsffile{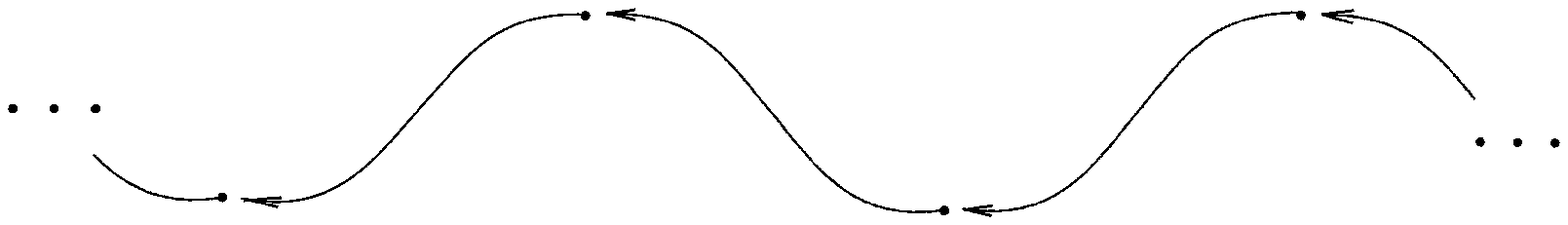}
\caption{\footnotesize {\it Jumping recombinant phase}: a {\it chain}
is naturally formed.\hspace{7.5cm}}
\end{figure}

%\bigskip

%\hrule height 0.4mm

%\bigskip

What one expects is that in the dense
instanton gas (or liquid) recombination takes place between
monopoles and antimonopoles from {\it different} pairs, thus {\it picking up}
a {\it chain} of instantons from the liquid (see Figure 2).

The spaghetti vacuum pattern implies that such chains are actually spread
out through the entire volume and form a ``percolating cluster'' \cite{HT,pc}.

As in the $2+1$-dimensional case, the electric fields with
non-vanishing vorticities, caused by the moving monopoles and
antimonopoles, contribute to the Wilson averages in $3+1$ dimensions and
give rise to the area laws.

At the moment there is no absolutely convincing {\it theoretical}
argument in favour to this kind of
ideas; instead they obtained considerable support from
computer {\it experiments}.

\subsection*{``Experimental'' lattice results}

Lattice computer simulations are primarily
targeted at producing qualitative results in the spirit of (i) and
thus at providing a proof that the Yang-Mills functional integral
indeed describes a theory with a mass gap, a linear potential,
a realistic hadronic spectrum and realistic hadron interactions.
Remarkably enough, these experiments could also be used for research
in direction (ii) and they indeed produced very inspiring
results. However, up to now the simulations are not
very detailed and one actually substitutes the functional integral
by a sum over a rather small random subset of field configurations,
which are believd to give the dominant contribution.
According to (ii) one can hope
that most of these dominant configurations will have something in
common -- and this is what really happens -- providing a clear description
of the {\it medium} required in (ii).

This {\it experimentally} discovered \cite{green,Amb} medium appears
to be somewhat unexpected (see \cite{orNO} for the original suggestion
of this ``Copenhagen spaghetti vacuum'' and
\cite{confmed} for comprehensive modern reviews and references):
it turns out to be filled with peculiar {\it one-dimensional} objects
(with two-dimensional world surfaces) -- $P$-vortices -- which in a certain
Abelian approximation (see next subsection)
look like narrow (of width
$r_m \ll \Lambda_{QCD}^{-1}$) tubes of {\it magnetic} field,
directed along the tube and changing direction to the opposite at
locations of monopoles and antimonopoles, which
form a 1-dimensional gas inside the tube
\footnote{
In contrast to the $P$-vortices themselves, the
monopoles and antimonopoles inside them are difficult to
define in a gauge-invariant way. Even the direction of
the would be Abelian magnetic field and thus
the exact positions of monopoles
and antimonopoles inside the $P$-vortex are unphysical: they can
be changed by gauge transformations.
Indeed, to change the {\it direction} of an Abelian field strength
$F^3_{\mu\nu}$ at a given point it is enough to make a singular gauge
transformation, conjugating the fields by a unitary matrix like $\sigma^1$
at this point (though it is not absolutely clear how to make such operation
consistent with the maximal Abelian projection, described in the
next subsection).
There is still controversy in the literature
(see, for example, \cite{KLM} for different points of view) about the
actual internal structure of the $P$-vortices and the (dis)advantages
of visualizing it in terms of monopoles and antimonopoles.
}.
Such objects are obviously
stable against the creation of monopole-antimonopole pairs: such processes
can not break the tube into two, because the magnetic flux through any section
outside the monopole cores is $1/2$
\footnote{
This does not contradict
the possibility that isolated monopoles are screened \cite{val}.
}.
The net of these direction-changing color-magnetic tubes
fills the entire space \cite{HT} (forming a ``percolating cluster'' \cite{pc})
\footnote{
In addition to the precolating cluster, there exists also a
variety of non-percolating ones, also populated by monopoles. There is no
agreement in the literature on whether these non-percolating clusters are
lattice UV-artifacts or they actually contribute in the continuum limit.
},
%In
%accordance with \cite{Zakharov}, there exist IR ("percolating cluster") and
%UV $P$-vortices settled with IR and UV monopoles. It remains unclear to what extent
%these UV objects to be incorporated into the ultimate
%continuum picture of the confinement, see also \cite{fine}.}
and in this medium
the force lines of color-electric fields (emitted by sources of non-vanishing
$N$-ality) also form tubes (of width
$r_e \sim \Lambda_{QCD}^{-1}$), thus giving rise the to confinement phenomenon.
In lattice {\it experiments} the area laws for approriate Wilson-loop averages
are explicitly checked and the $P$-vortices from percolating cluster
are shown to give
dominant contribution to the string tensions.
{\it Theoretically}, the contribution of $P$-vortices to
the string tension depends on their abundance and one of
the tasks of the theory is to explain the origin of the medium of $P$-vortices
and how it is consistent with Lorentz invariance.
%\footnote{
%Contribution of $P$-vortices is much easier to compute for the
%averages of the {\it space} Wilson lines, which measure the flux
%of magnetic field instead of the line integral of electric field. Given Lorentz
%invariance, the area laws for the space loops implies that for the
%time ones.
%}

So far there is no clear theoretical explanation of why and how such
a medium is formed in non-Abelian gauge theories and why -- once formed --
it can give rise to a dual Meissner effect and lead to confinement, though
the (lattice) experimental evidence in favour of this pattern is
rapidly growing.

A serious drawback of the published results of lattice experiments is
that they do not provide the essential information about instanton-like
and meron-like configurations and their probable association with the
{\it localized} $P$-vortex clusters and, furthermore, they do not explicitly
study the configurations of collimated color-electric force lines between
sources with non-vanishing $N$-ality (which do not need to be fermions).
Information about these color-electric tubes
is extracted indirectly from the study of Wilson averages. This is not
enough to understand what happens to these tubes,
say, after the maximal Abelian projection,
and whether their content indeed looks like Abelian electric field
exactly in the same projection, when the $P$-vortices look like the tubes of
direction-changing Abelian magnetic field.
Any data touching upon this issue would be
extremely useful for further clarification of the situation.

\subsection*{Maximal Abelian Projection}

The ``$P$'' in ``$P$-vortices'' comes from the word ``projection'' \cite{Pvor}.
It is inspired from the way they are often searched for and studied,
which is not gauge invariant,
even though the $P$-vortices themselves are in fact gauge invariant
(see Figure 3).

Usually, one uses a procedure called the Maximal Abelian Projection
(MAP)
\footnote{
Comparison with the results of lattice experiments in other Abelian
approximations usually demonstrates that the (gauge non-invariant and
necessarilly approximate) language of monopoles is most reliable in
the MAP, the use of this language in other calculational schemes can often
be misleading \cite{chern}.
}.
It splits into two steps. First,
for every configuration of the fields $A^a_\mu(x)$, taken with the
weight dictated by the true non-Abelian action,
the ``maximal Abelian gauge'' is chosen, by minimizing the lattice
counterpart of $\int W^+_\mu W^-_\mu(x) d^4x $ along the gauge orbit.
This first step is absolutely justified (though technically it suffers
from ambiguities caused by the existence of Gribov copies).

%\bigskip

%\hrule \hrule height 0.4mm

%\bigskip

\paragraph{\large {\bf Figure 3.}}

{\footnotesize
This figure, borrowed from the seminal paper \cite{Amb}, is the best
existing illustration of what $P$-vortices are and what the maximal
Abelian projection does.}

\bigskip
\setcounter{figure}{0}
\def\thefigure{\alph{figure})}
\begin{figure}[h]
\begin{center}
\epsfxsize 300pt
\epsffile{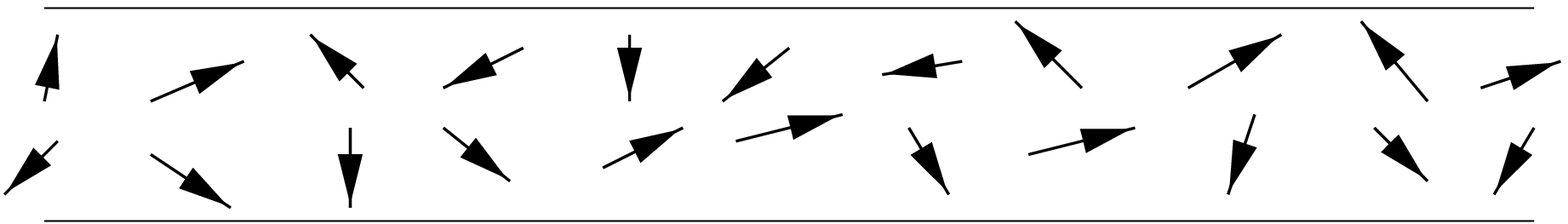}
\caption{\footnotesize
A fragment of the distribution of field strength in an original
configuration of
fields $A^a_\mu(x)$, from the set of those which give dominant
contribution to the non-Abelian functional integral.  The strenghts are
non-vanishing within a narrow tube, the $P$-vortex. Actually, the entire
configuration looks like a net of $P$-vortices, containing the
``percolating cluster'', which has proper scaling properties and survives
in the continuum limit. The arrows indicate directions in {\it color} space.}
\end{center}

\bigskip

\begin{center}
\epsfxsize 300pt
\epsffile{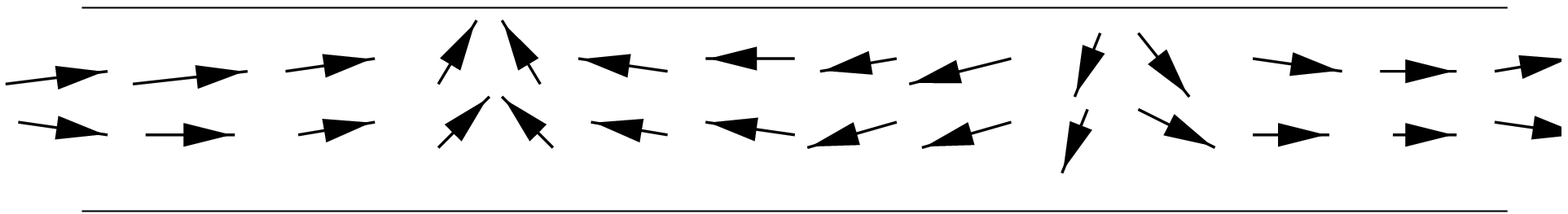}
\caption{\footnotesize
The maximal Abelian gauge is chosen, which minimizes
$\int W^+_\mu W^-_\mu(x) d^4x$. It is just a choice of gauge
(field strenths are rotated), no approximation is
involved. Certain structures are clearly seen in the distribution
of field strenghts {\it inside} the tube.}
\end{center}

\bigskip

\begin{center}
\epsfxsize 300pt
\epsffile{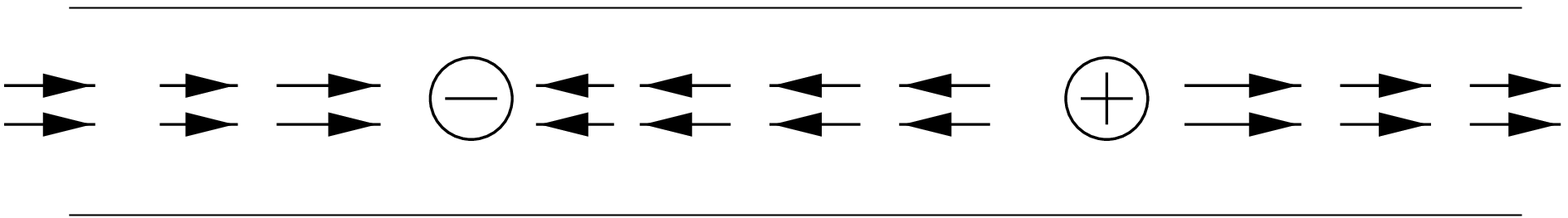}
\caption{\footnotesize
Maximal Abelian {\it projection} is performed: $W^\pm_\mu(x)$ are put
equal to zero. The {\it structures} seen in (b), turn into a clear
(but approximate) pattern of collimated magnetic force lines, changing
direction at the location of monopoles and antimonopoles. No peaks of
magnetic energy occur at these locations.}
\end{center}
\end{figure}

%\bigskip

%\hrule \hrule height 0.4mm

%\bigskip

This allows one to introduce the induced effective action
$\tilde S(A)$, obtained after
integration over the other components ($W^{\pm}_\mu\equiv A_{\mu}^1\pm iA_{\mu}^2$,
$D_{\mu}(A)\equiv \partial_{\mu}+ieA_{\mu}^3$)
\be
\exp \Bigl(-\tilde S(A)\Bigr) =
\int
DW^+ DW^-
\ \delta\left(\left|D_\mu(A) W^+_\mu\right|^2\right)
%\delta\left(D_\mu(A) W^-_\mu\right)
\hbox{det}_{FP}^2 \left(\partial_\mu D_\mu(A)\right) \times \\\times
\exp -\frac{1}{g^2}\left[
\left(
F_{\mu\nu} + (W^+_\mu W^-_\nu - W^-_\mu W^+_\nu)
\right)^2
+ |D_\mu(A) W^+_\nu|^2
\right]  \ \
\label{SU2MAP}
\ee

At the second step, one makes use of $\tilde S(A)$ to define Abelian
correlation functions
\be\label{17}
\langle \prod_i {\cal O}(A^a_\mu) \rangle_{MAP} \equiv
\langle \prod_i {\cal O}(W^\pm_\mu = 0, A^3_\mu) \rangle \ =
\int DA^3_\mu e^{-\tilde S(A^3_\mu)}
\prod_i {\cal O}(W^\pm_\mu = 0, A^3_\mu)
\ee
This step implies that one makes use of the true non-Abelian action,
i.e. includes contributions from the {\it virtual} $W^\pm$-bosons
in loops, but omit them from external lines.
Therefore, the second step -- the {\it projection} itself --
is an approximation:
\be
\langle \prod_i {\cal O}(W^\pm_\mu, A^3_\mu) \rangle \ne
\int DA^3_\mu e^{-\tilde S(A^3_\mu)}
\prod_i {\cal O}(W^\pm_\mu = 0, A^3_\mu)
%\hbox{Equation (\ref{17})}\ne \langle \prod_i {\cal O}(W^\pm_\mu, A^3_\mu) \rangle
\ee
Its experimentally discovered \cite{Suzuki} surprising efficiency
(as compared with the complete answer including non-Abelian
fields) is often called the {\it hypothesis
of Abelian dominance}. Though theoretically so far unjustified and
uncontrollable, it provides a convenient {\it language} for
description (visualization) of the confinement phase: it is at this level
that monopoles and antimonopoles appear. Figure 3 can serve as an
illustration of how the MAP works.

The theoretical problem of evaluation of $\tilde S(A)$ remains open.
See \cite{FN} for interesting attempts to identify condensating modes
and vortex-like structures in the functional integral (\ref{SU2MAP})
and \cite{Tong} for a supersymmetric model with BPS configurations, which
look like magnetic P-vortices populated by monopoles.

\section{Are there condensed-matter analogies of confinement?}

Coming back to the lattice results above,
a natural question to ask is if anything similar can be found
in other avatars of gauge theories, for example, in condensed matter
or plasma physics. There, one would rather expect to encounter a dual
type of medium: electric $P$-vortices, formed by chains of
positive and negative electric charges, connected by narrow tubes
of electric fields with fluxes $\pm 1/2$, {\it and} an ordinary (magnetic)
MA effect, implying formation of magnetic-field tubes with a constant
unit flux (and confinement of hypothetical magnetic charges), {\it caused by}
or at least {\it consistent with} the existence  of such electric
$P$-vortices. In condensed matter analogues, the underlying non-Abelian
Yang-Mills dynamics responsible
for the formation of $P$-vortices, should
presumably be replaced by some other dynamics (additional forces),
allowed in condensed matter systems. The whole situation (the coexistence and
even mutual influence of electric $P$-vortices and magnetic MA effect) is
already exotic enough to make one wonder if anything like this can at all
occur in any kind of natural matter systems.

The main thing to look for in a condensed matter setup is the
{\bf simultaneous existence of narrow
tubes ($P$-vortices) of direction-changing electric field and broader tubes
(Abrikosov lines) of magnetic field}
-- a dual pattern to the one,
underlying the spaghetti confinement mechanism of gluodynamics.
This clearly implies,
that superconductivity (from the dual superconductor scenario),
if relevant at all, should be of a more sophisticated nature than just
the single-field condensation (monopole condensation), the
superconducting order should be caused or at least coexist with
an order of some other type (responsible for the formation of $P$-vortices).
This looks almost like the requirement that the
Meissner-Abrikosov effect (for magnetic
field) {\it coexists} with (or, perhaps, is even {\it implied by})
the dual Meissner-Abrikosov effect
(for electric field), but actually the tubes of electric field should be
different: they should have internal structure, namely the one-dimensional
gas of positive and negative electric charges, an electric field along the tube
which changes direction at the locations of these charges and be
stable against possible ``string breaking'', caused by creation or annihilation
of charge-hole pairs. Moreover, the width of electric tubes should/can
be different (much smaller?) than that of magnetic tubes.

The main goal of this paper is to bring these issues to the attention of
experts in other fields, such as
condensed matter and plasma physics and to emphasize the fact
that the discovery of a similar picture
arising under any circumstances, would be of great help
for the development of the confinement theory and in particular for the
understanding of possible $2d$ vortex theories, living on the world
sheets of the relevant branes, as well as of the phase structure of
these theories
\footnote{
Among other things, it would be interesting to exploit the idea of
the {\it topological} confinement which, in different versions, often works
in condensed matter physics. A characteristic feature of the topological
confinement is that it depends on the dynamics of the theory only
through the properties of particular excitations (quasiparticles),
while their interactions do not matter. For example, one-dimensional
objects can be tied and, therefore, be unseparable, and this can work
for real one-dimensional excitations, like Abrikosov tubes, and for
point-like magnetic monopoles and/or hedgehogs, which have attached Dirac
strings. In practice,
topological confinement can look very similar to the mechanism we
discuss throughout the paper. See \cite{Vol,Volt} for some examples, see
also \cite{J}.
}.
If, on the contrary, no such pattern exists in condensed matter physics,
this would once again emphasize the pecularities of non-Abelian gauge theories
(where elementary quanta carry more structure than just point-like charges
and thus the naive screening behaviour is from the very beginning
substituted by antiscreening and further non-naive phenomena are
naturally expected to occur).

The rest of this paper is purely speculative, added for encouragement:
in order to demonstrate that superconductivity (probably responsible
for the magnetic Meissner-Abrikosov effect) can indeed {\it coexist} with
at least some kind of {\it dual} order (though the example below falls
short from exhibiting narrow tubes of direction-changing electric field).

\subsection*{Charge density waves}

As a possible (but by no means the only) candidate analogue of the
electric $P$-vortices we would like to suggest the
{\it charge density waves} (CDW) and the questions which arise are:

(a) are there are any {\it tube-like} CDW
with a charge density similar to $\rho(x,y,z) \sim \delta^{(2)}(x,y)\sin z$
and (perhaps, direction-changing) electric force lines collimated along
the $z$ axis?

(b) can the CDW {\it coexist} with superconductivity (SC),
which would be a natural reason for the Meissner-Abrikosov effect?

(c) can the CDW {\it cause} or at least {\it enhance}
superconductivity?

(d) can the widths of the CDW-like $P$-vortices be much smaller
than those of Abrikosov lines (where the Cooper-Higgs-like condensate
is broken)?

Remarkably, a very similar set of questions is currently under intense
investigation in connection with high-$T_c$ superconductivity
(where the adequate theoretical pattern also remains unknown),
and it looks like the above possibilities are indeed open, as one may see
in \cite{CDWrev} and references therein.
Of course, the real media appearing in condensed matter examples,
have a lot of additional structure (primarily, the highly anisotropic
crystal lattice in the background, playing a key role in
the formation of realistic CDW), which one does not expect to find
in gluodynamics. For closer analogies with gluodynamics one
can also look for phenomena in liquid ${He}$ \cite{Vol}, dense relativistic
plasma, segnetoelectrics \cite{Kir} or even biological membranes \cite{nelson}.
Still, we want to emphasize once again that today, when the formulation
of a {\it phenomenological theory} of $P$-vortices is so important,
one needs to consider {\it all} examples, where objects
of this kind are presumably present, irrespective of the underlying
microscopic structure, and the solid-state systems with the coexisting
CDW and SC orders should not be neglected -- especially because, like
the confinement in gauge theories, they are now under close scrunity, and
considerable progress can result rather fast from comparison of ideas
from the two fields.

The simplest facts and ideas about the CDW-SC systems, though not
immediately coinciding with (a)-(d), do not seem to be in obvious
contradiction also. The relevant list of properties seem to include
the following:

$\bullet$ The CDW formation causes transition to an insulator phase
(Peierls-Frohlich-Mott transition), while SC transition gives rise to a
(super)conductor.

$\bullet$ Thus CDW and SC orders compete with each other, with CDW
usually stronger competitor than SC \cite{AFK}.

$\bullet$ Still the CDW and SC orders can coexist \cite{coex,Lee}.

$\bullet$ Even if both CDW and SC orders are not established simultaneously
at long distances, they interfere locally, one phase appears in the regions
where the other is broken:
SC appears in the vicinity of CDW vortices and
CDW appear in the vicinity of Abrikosov lines \cite{Lee}.
This can be enough, for example, to get the SC phase in when CDW
dislocations percolate through the entire volume.

The phenomenological descriprion of the CDW is in terms of electron-phonon
interactions \cite{ME}. Notice that the vector nature of
phonons makes them closer to the $W$-fields in (\ref{SU2MAP}) than the
scalar fields, employed in the Abelian Higgs model (\ref{ah}).

\section{Conclusion}

The {\it theory} of the {\it Copenhagen spaghetti vacuum} should, of course,
be developed in the context of string theory. The appropriate name
for $P$-vortices is $1$-branes. Monopoles living on these $1$-branes are,
naturally, $0$-branes. The coexistence of electric and magnetic Abrikosov
tubes should be modelled by that of coexisting ``fundamental strings'' and $D1$
branes. The problems, raised in this paper, are related to the
lack of any ``underlying model'', for which the theory of strings and branes
would be an effective model, the lack which seriously undermines the
progress in modern string theory.
We emphasize that the spaghetti vacuum in gluodynamics can by itself provide
such a model and we also suggest to start a more
extensive search for possible underlying models in modern condensed-matter
physics.

\section{Acknowledgements}

We are grateful to T.Mironova for help in making figures.
This work was supported in part by the EU under the RTN contract
MRTN-CT-2004-512194.
The A.M.s acknowledge the support of two NATO travel grants and
the hospitality of the Department of Physics of the University of Crete, where
this work was done.
This work was also partially supported by the Federal Program
of the Russian Ministry of
Industry, Science and Technology No 40.052.1.1.1112 and
by Volkswagen Stiftung, by the grants
INTAS 00-561, RFBR 04-02-16538a (Mironov),
INTAS 00-561, RFBR 01-02-17488 (Morozov) and by
the Grant of Support for the Scientific
Schools 96-15-96798 (Mironov).

\end{document}